\documentclass[epj,referee]{svjour}
\usepackage[pdftex]{graphicx}

\begin{document}
\title{Why Y chromosome is shorter and women live longer? }
\author{Przemyslaw Biecek\inst{1} \and Stanislaw Cebrat\inst{2}}

\institute{${}^1$przemyslaw.biecek$@$gmail.com, ${}^2$cebrat$@$smorfland.uni.wroc.pl, Department of Genomics, University of Wroclaw, ul. Przybyszewskiego 63/77, 51-148 Wroclaw, Poland, phone: +48-71-3756-303.}
\date{Received: date / Revised version: date}
%
\abstract{
We have used the Penna ageing model to analyze how the differences in evolution of sex chromosomes depend on the strategy of reproduction. In panmictic populations, when females (XX) can freely choose the male partner (XY) for reproduction from the whole population, the Y chromosome accumulates defects and eventually the only information it brings is a male sex determination. As a result of shrinking Y chromosome the males become hemizygous in respect to the X chromosome content and are characterized by higher mortality, observed also in the human populations. If it is assumed in the model that the presence of the male is indispensable at least during the pregnancy of his female partner and he cannot be seduced by another female at least during the one reproduction cycle - the Y chromosome preserves its content, does not shrink and the lifespan of females and males is the same. Thus, Y chromosome shrinks not because of existing in one copy, without the possibility of recombination, but because it stays under weaker selection pressure; in panmictic populations without the necessity of being faithful, a considerable fraction of males is dispensable and they can be eliminated from the population without reducing its reproduction potential. 
} 
\maketitle
\section{Introduction}
\label{intro}
Sexually reproducing diploid organisms possess two sets of chromosomes, (in case of humans each set is composed of 23 chromosomes). In mammals, females' genomes consist of fully homologous sets while males have different chromosomes in one pair. This pair is called sex chromosomes; one chromosome is called Y and the second one X. Females possess two X chromosomes, produce haploid gametes each with the X chromosome that is why this sex is called homogametic. Males produce half of gametes with the X chromosomes and half with the Y chromosome and are called heterogametic sex. In reptiles or birds the male sex which is homogametic and the  female sex is heterogametic. To underline the difference in homo-/heterogametic sexes geneticists introduced different connotations for sex chromosomes: XY for species with heterogametic males and ZW for species with heterogametic females. Ohno \cite{Ohno}, basing on his observations concluded that both sex chromosomes in reptiles originated from the same pair of ancestral chromosomes. This conclusion was adopted also for mammalian sex chromosomes though, it has been assumed that the mammalian sex chromosomes evolved from a different pair. Vertebrate males share the same system of sex determination, the same cascade of genes responsible for differentiation of testes and similar spermatogenesis processes (see \cite{Ferguson} for review). Mammalian Y chromosomes, similarly to W chromosomes in reptiles, show significant variation in size between species but it always contains genes responsible for male sex determination. There is a still debated problem: why Y chromosome in mammals was reduced in size during the evolution. Some authors suggest that the main reason for genetic shrinking of Y was switching off the recombination between X and Y chromosomes. After the recombination stops, the Muller ratchet mechanism allows the mutation accumulation in the clonally propagating chromosome. Lobo and Onody \cite{Lobo} have shown in the computer modeling that the Y chromosome degenerates even if both pairs XX and XY do not recombine. Authors try to explain these phenomena by the specific differences in inheritance of the sex chromosomes - the Y chromosome never experiences the selection pressure in the female body while X chromosomes spent one third of their evolution time in the male body and two thirds in the female body. Nevertheless, we have proven in our computer modeling that switching off the recombination between sex chromosomes is not enough to induce the degeneration of the Y chromosome. It is rather a specific strategy of reproduction which is responsible for this process. 

\section{Model} 
Our model is based on the  diploid sexual Penna model of ageing \cite{Penna} see \cite{Stauffer} for review. 
In the model, each individual is represented by two pairs of homologous chromosomes: one pair of sex chromosomes and one pair of autosomes (chromosomes other than sex chromosomes). Each chromosome is represented by a vector of $L=32$ bits 0's or 1's. Thus the genotype of an individual $i$ is represented by $4*L$ digits. The genome of individual $i$ is denoted as
$$
G_i = \{g_{i,j,k} \in \{0,1\}, j\in \{1, ..., 4\}, k\in\{1, ..., L\}\}.
$$

In our simulations we consider two kinds of sexual systems. The first one, denoted \textbf{XX/XY}, corresponds to the populations in which females have two homologous X chromosomes, while males have two different sex chromosomes: X and Y. The second system, denoted \textbf{ZZ/WZ}, corresponds to the populations in which males have two homologous Z chromosomes while females have two different sex chromosomes: Z and W. The first system is typical for mammals, the second one for birds but there are also some other animals which exploit it in the sex determination.

In the model we have introduced one recombination between homologous pair of chromosomes during the gamete production, i.e. recombination is allowed between autosomal chromosomes, and flso in the emale XX pair in the first system or male ZZ pair in the second system.

We consider two reproduction strategies. In the first strategy, called ,,random mating'', every year (Monte Carlo step - MCs) the female at the reproduction age (at least 8 years old; $R=8$) chooses randomly a male partner. In the second strategy, called ,,faithful to the death'', the female at the first year of the reproduction age looks for a partner at the reproduction age and creates with him a couple for the rest of their life. She will not mate with other partners. If the partner is not available she will wait and try to find him in the following years, when other males reache the minimum reproduction age.

During reproduction, the parental genomes are replicated. Into each pair of chromosomes one mutation is randomly introduced (every gene is mutated with probability $1/{2L}$). If the value of the mutated gene is 0 it is changed to 1, otherwise it stays 1. Next, one recombination between autosomes and one between homologous pair of sex chromosomes occur in a randomly chosen point and one chromosome of each pair is chosen to form the gamete. One gamete from each parent is transferred to the offspring genome.  The newborn survives the first year with a probability given by Verhulst coefficient
$$
V_{survive} = 1 - \frac{N_{act}}{N_{max}},
$$
where $N_{max}=2000$ is the maximum capacity of the environment while $N_{act}$ is the actual population size.

The alleles are switched on chronologically during the whole lifespan of individual; in the first year two alleles in the first locus of autosomes and two alleles of the first locus on sex chromosomes are activated. In the second year the alleles of the second loci of both pairs are activated and so on. The individual dies in age $a(i)$ when $T=3$ pairs of defective alleles are switched on, which means that on the positions till $a(i)$ the number of paired defects on autosomes and sexual chromosomes is $3$	.
In other words, all mutations are recessive and individual $i$ dies when
$$
\sum_{k=1}^{age(i)} g_{i,1,k} \cdot g_{i,2,k} + g_{i,3,k} \cdot g_{i,4,k} \geq T.
$$
where $age(i)$ is the age of individual $i$. The live span $a(i)$ is the minimum $age(i)$ for which the above condition is satisfied.

Simulations start with all alleles set to 0, i.e. with individuals with the perfect genomes, without any defect.

\section{Results} 

\subsection{XX/XY system - panmictic population}
As it was described in the above section, the individuals are represented by two pairs of chromosomes - one pair of autosomes and one pair of sex chromosomes. At the beginning of simulation all four chromosomes are identical, genetically perfect but in the male genome one of the sex chromosomes is marked as Y. Each newborn inheriting this chromosome from its father will be a male, otherwise a female. In the panmictic population each male after reproducing with one female returns to the total pool of males and can participate during the same MCs (year) in the reproduction process with another female. In Fig. 1 we have shown the distribution of defective alleles along the autosomes, and the sex chromosomes. 

\begin{figure}
\resizebox{0.95\columnwidth}{!}{%
		\includegraphics[width=0.8\textwidth]{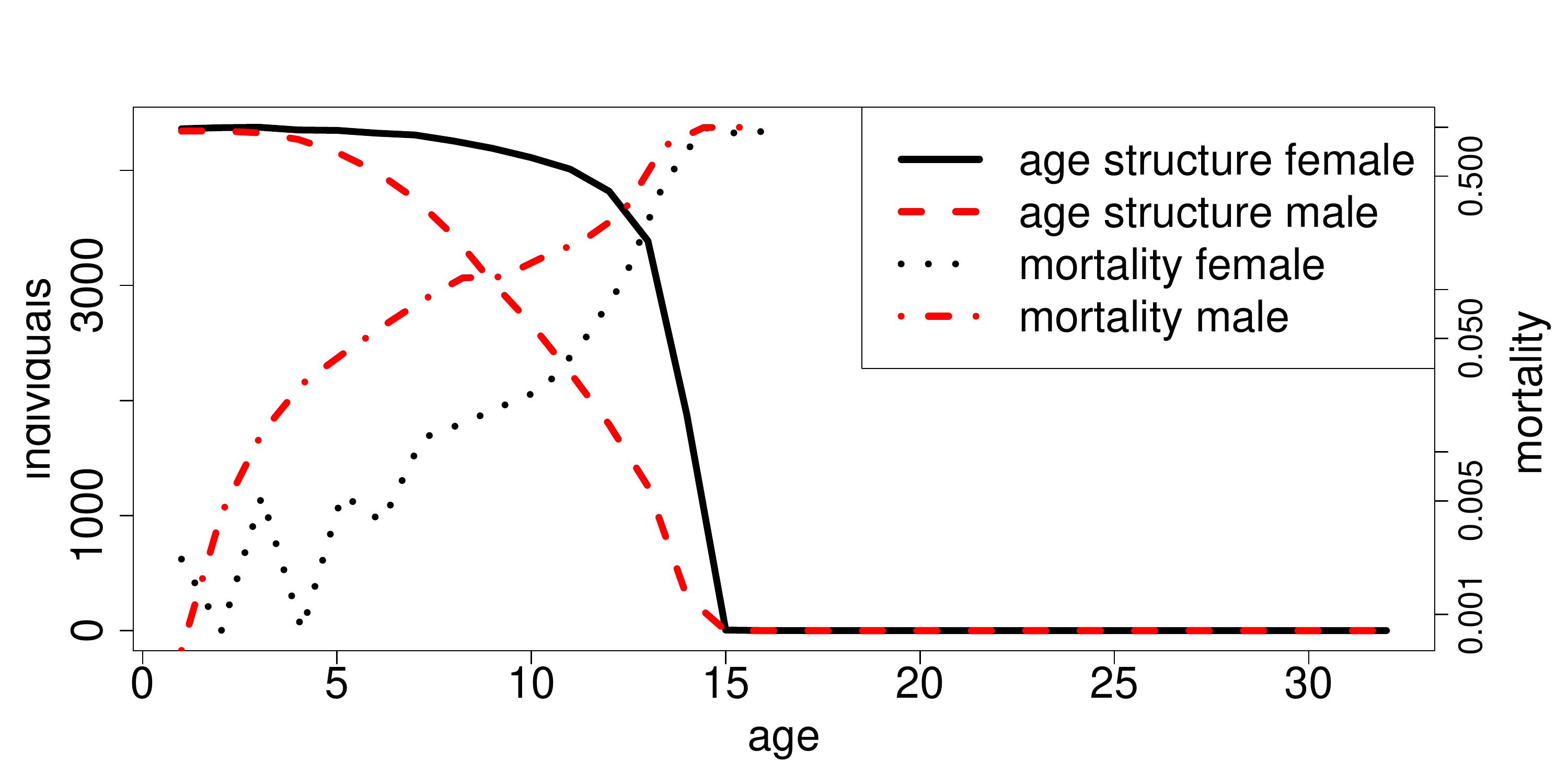}
}
\resizebox{0.95\columnwidth}{!}{%
		\includegraphics[width=0.8\textwidth]{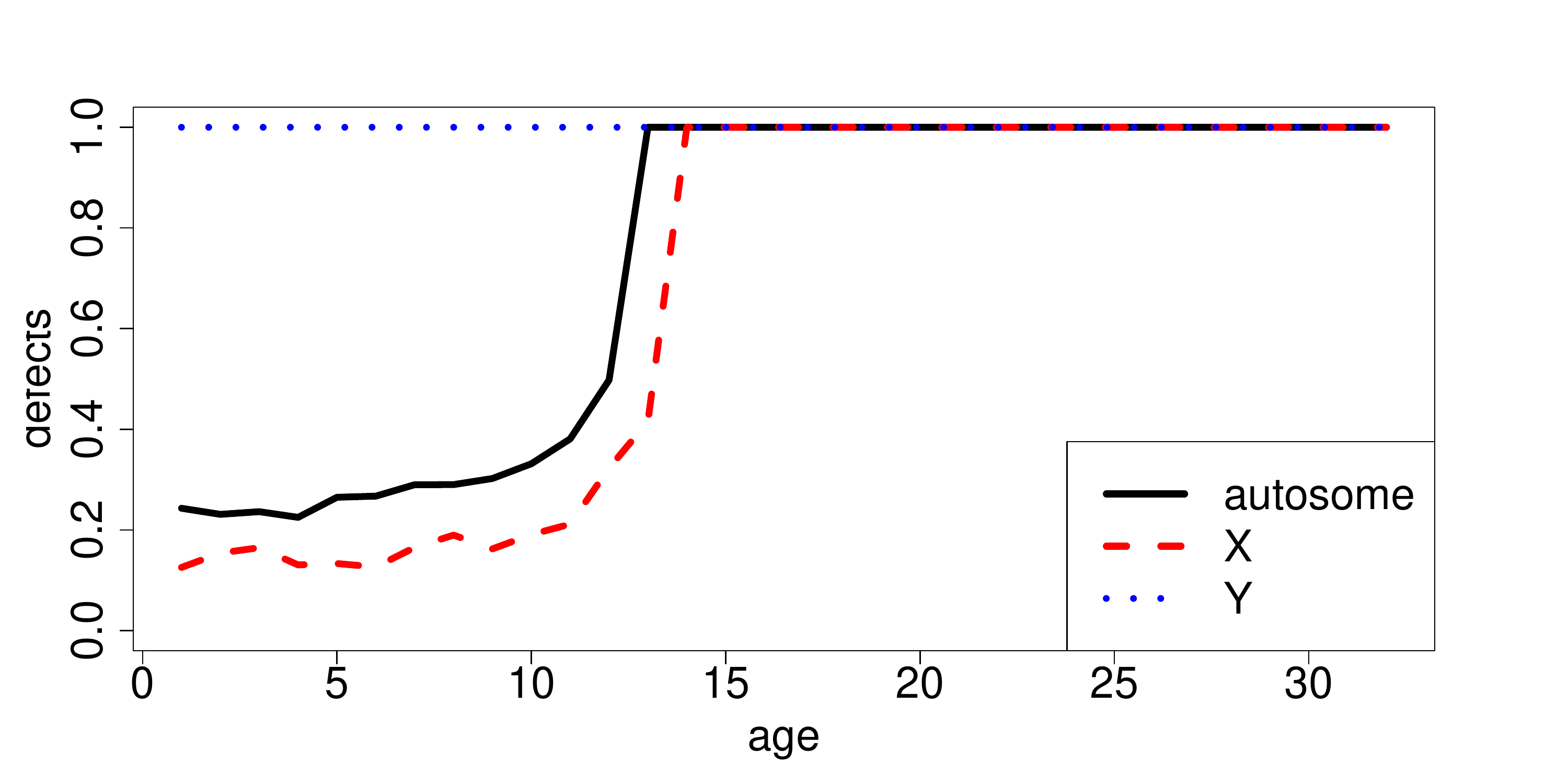}
}
	\caption{XX/XY system - panmictic population. Age structure and mortality in populations after 20 000 MCs (the upper plots). Note the logarithmic scale (right) for mortality. Lower plots represent fraction of defective alleles in autosomes and sex chromosomes.}
\label{fig:2}       
\end{figure}

\begin{figure}
\resizebox{0.95\columnwidth}{!}{%
		\includegraphics[width=0.8\textwidth]{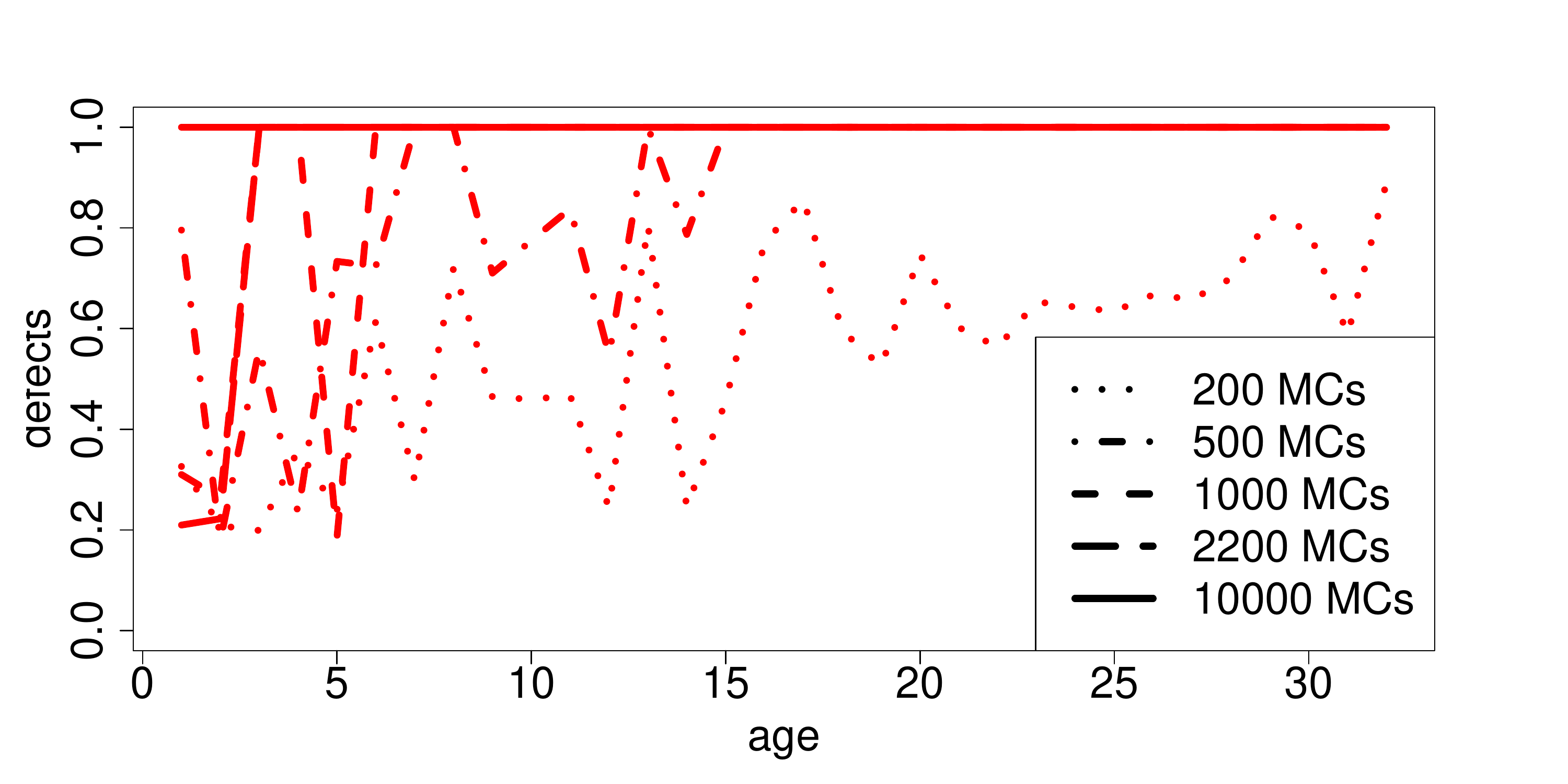}
}
\resizebox{0.95\columnwidth}{!}{%
		\includegraphics[width=0.8\textwidth]{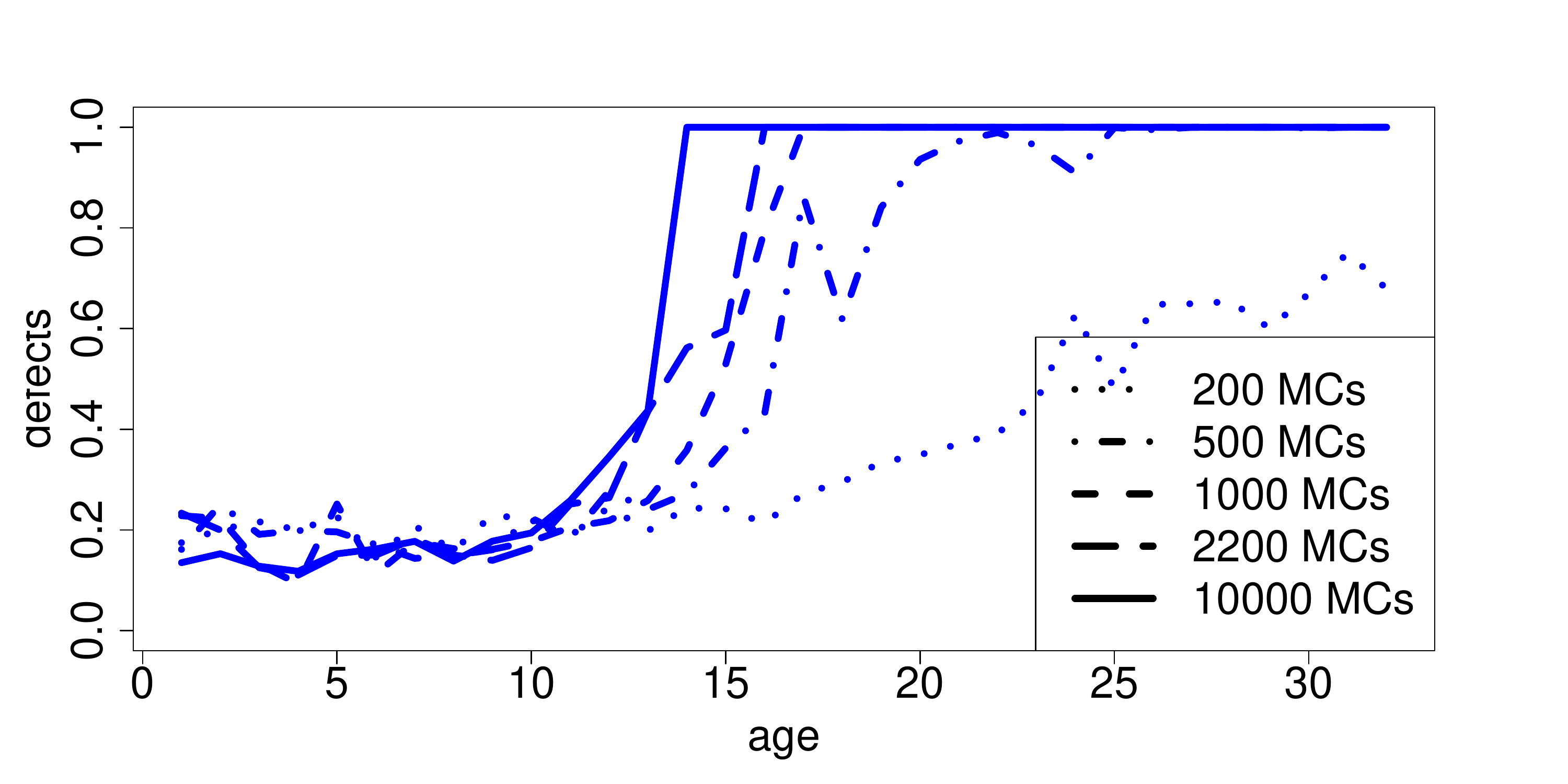}
}
	\caption{Evolution of chromosomes in the XX/XY system - panmictic population. Fractions of defective alleles along the Y chromosome (upper plots) and along the X chromosomes (lower plots) after different time of simulation.}
\label{fig:1}       
\end{figure}

In general, the Penna ageing model supports the assumptions of the Medaware hypothesis of defective genes' accumulation in the genetic pool of population as a main cause of ageing \cite{Medawar}. Genes expressed before the minimum reproduction age are under very strong selection pressure and the fraction of defective alleles in this part of genomes is low. Defects expressed after the minimum reproduction age could be transferred into the offspring with relatively higher probability and they are under weaker selection pressure. That is why we can observe the growing fraction of defects among these genes. The age, when the fraction reaches 1 it determines the maximum lifespan of individuals in the population. This specific distribution of defective alleles is observed in autosomes and X chromosomes after 20 000 MCs (population in equilibrium). All loci on the Y chromosome are already defective. Since we assume that all defective alleles are recessive, the defects on X chromosome could be complemented in the female genomes but are not complemented in the hemizygous genomes of males (the Y chromosome is already empty). That is why in the male genomes the defective genes on X chromosomes are under stronger selection than the defects on autosomes. As a result we observe a lower fraction of defective genes in the X chromosomes than in autosomes, what was also observed in the earlier simulations \cite{Kurdziel}. The lack of complementing the defective alleles located on X chromosomes in the male genomes affects the age distribution and mortality of males and females in the simulated populations (shown in Fig. 1, upper plots). At birth, the fractions of males and females are equal which is obvious under these parameters of simulations. However, the mortality of males, due to their role in the elimination of defects from X chromosome is higher. As a result, the total fraction of males in population is lower than females. The effect of higher mortality of men due to the hemizygozity in respect to genes located on X chromosome was also simulated previously \cite{Schneider}, \cite{Kurdziel} even if the number of genes located on sex chromosomes was considerably smaller. The  effect of degenerating Y chromosomes was simulated earlier by Lobo and Onody \cite{Lobo}. The different evolutionary history of sex chromosomes is shown in Fig. 2. The Y chromosomes accumulate defects expressed in the earlier ages of individuals from the very beginning of the simulations while the accumulation of defects in autosomes and X chromosomes is highly biased - a lower number of accumulated defective genes expressed before the reproduction age and a higher number for genes expressed later during the lifespan.

\subsection{XX/XY system - faithful pairs}
In the second version of simulations we have changed the model only in one point; the male, after reproduction process with one female, does not return to the total pool of males in the population. This version gives similar results to the simulations when males and females are ,,faithful to death'' which means that if a female at the reproduction age finds a male at the reproduction age they both stay in the same pair until the death and even the death of one partner does not allow the surviving one for looking for another partner. Now the evolution of sex chromosomes is significantly  different than in the previous version, as shown in Fig. 3. 

\begin{figure}
\resizebox{0.95\columnwidth}{!}{%
		\includegraphics[width=0.8\textwidth]{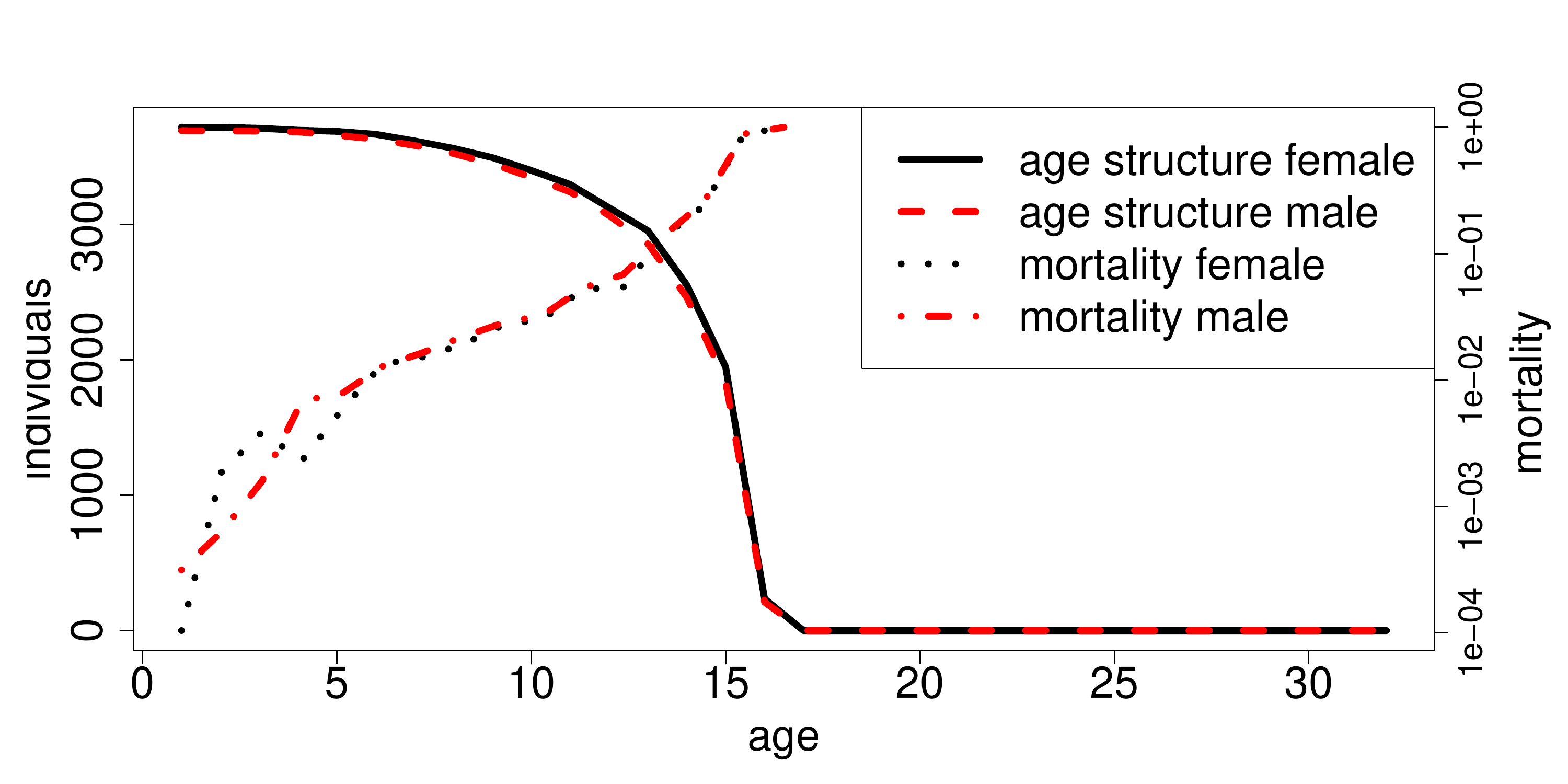}
}
\resizebox{0.95\columnwidth}{!}{%
		\includegraphics[width=0.8\textwidth]{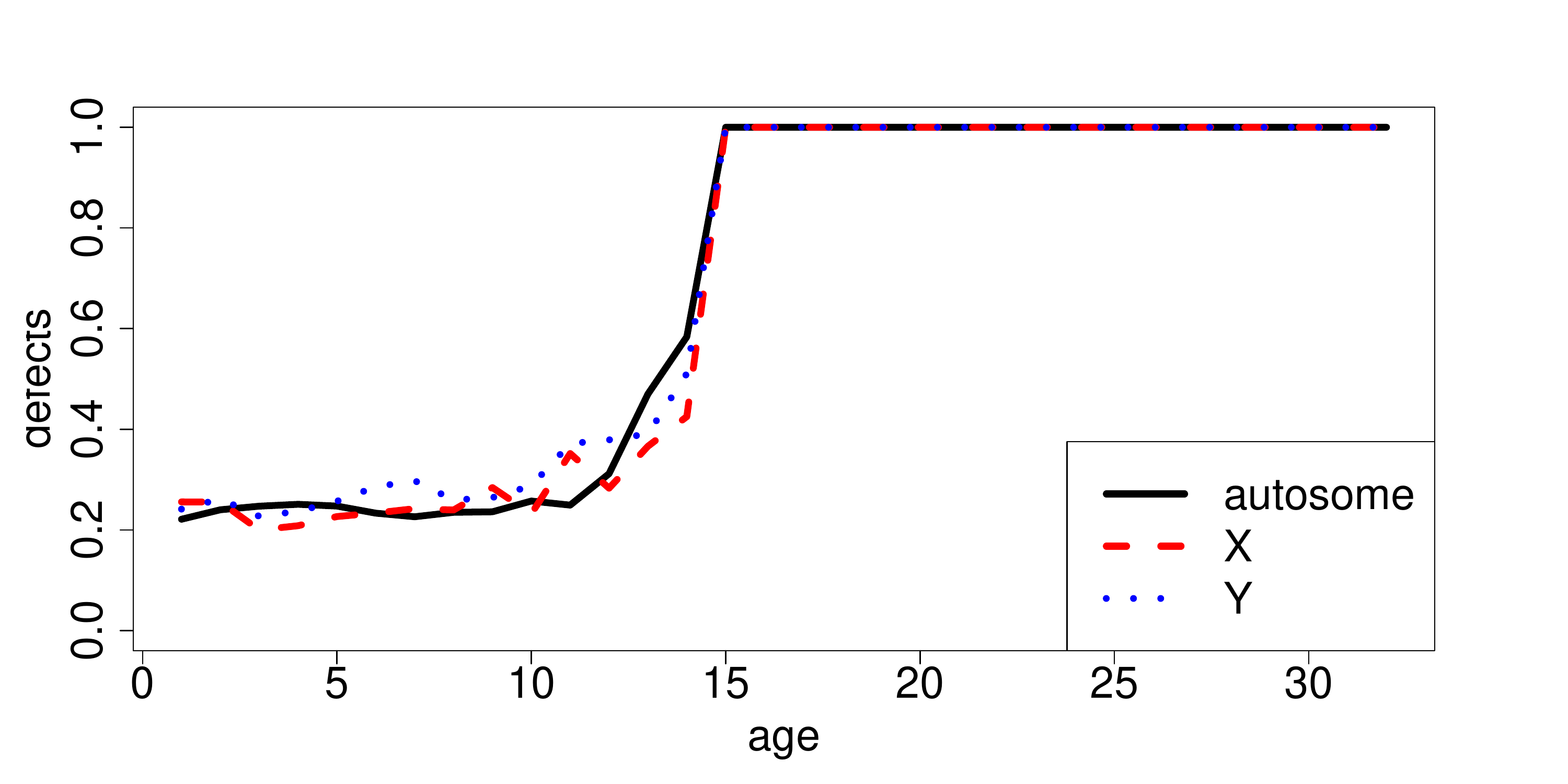}
}
	\caption{XX/XY system - faithful pairs. All descriptions like in Fig. 1.}
\label{fig:2}       
\end{figure}

The age distribution and the mortality of both sexes are similar. Also the distribution of defective genes along the autosomes and both sex chromosomes is the same. Comparing the results of the two simulations one can conclude that it is the difference in the strategy of reproduction which is responsible for the shrinking Y chromosome and differences in the mortality of males and females. In the first strategy, the reproduction potential of population depends on the fraction of females at the reproduction age in the population. The size of the fraction of males is unimportant for the reproduction potential or even a smaller fraction of males could be advantageous because they contribute to the total size of population controlled by the common Verhulst factor \cite{Verhulst}. In the second version, the fraction of both sexes are equally important for the reproduction potential of the population. That is why selection keeps the mortality of both, males and females at the same level and any loss of information from Y chromosome increases the male mortality and is disadvantageous for the reproduction potential of the population.

\subsection{ZZ/ZW system}
Evolution of sex chromosomes in the ZZ/ZW system is different. In this system, like in birds, males are homogametic and females are heterogametic. If the scenario of chromosome evolution is similar to the XY system then, females should be characterized by higher mortality and the reproduction potential of the whole population would decrease. Thus, selection has to keep the information of the W chromosome eliminating defects from it. 

\begin{figure}
\resizebox{0.95\columnwidth}{!}{%
		\includegraphics[width=0.8\textwidth]{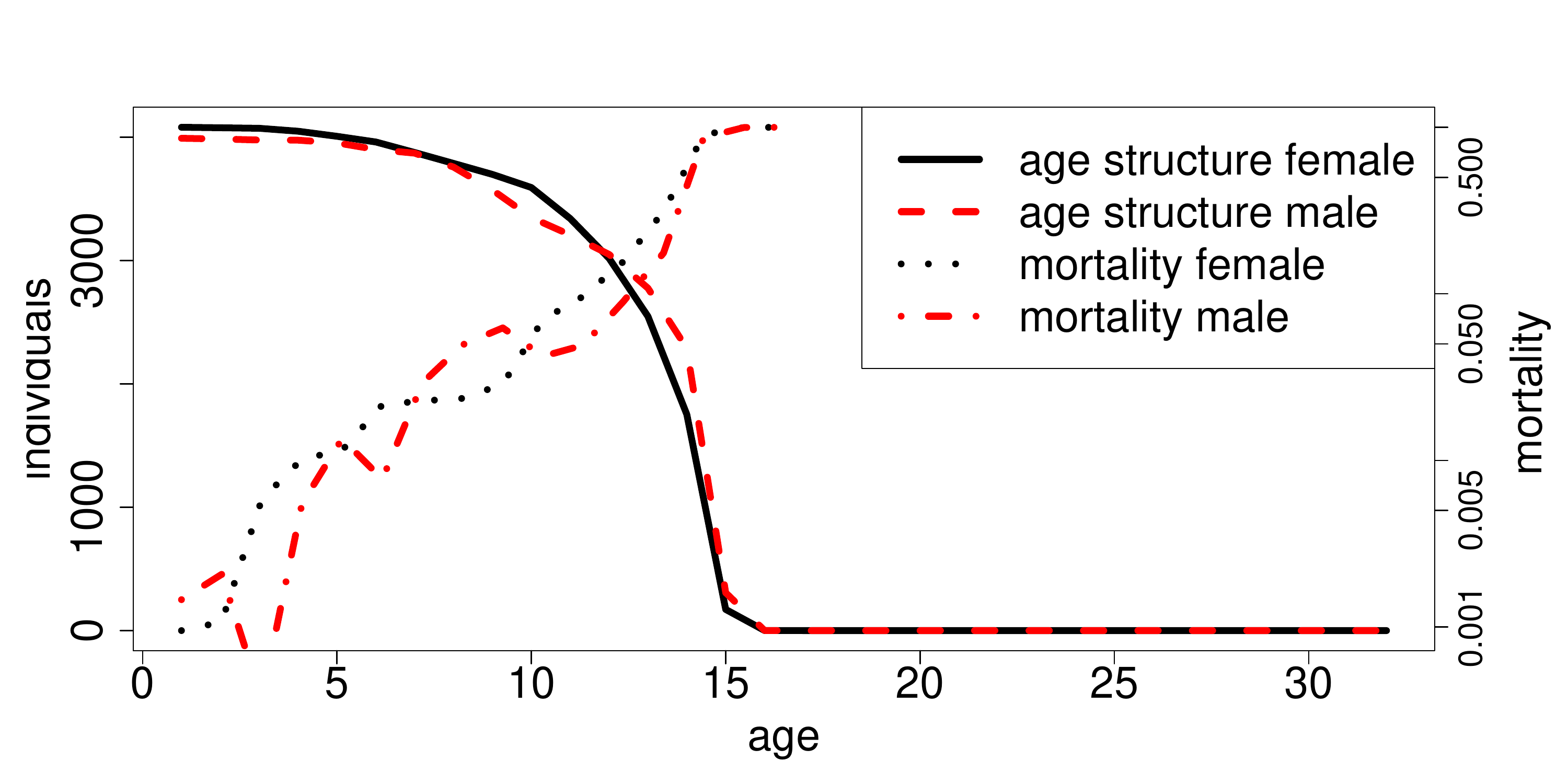}
}
\resizebox{0.95\columnwidth}{!}{%
		\includegraphics[width=0.8\textwidth]{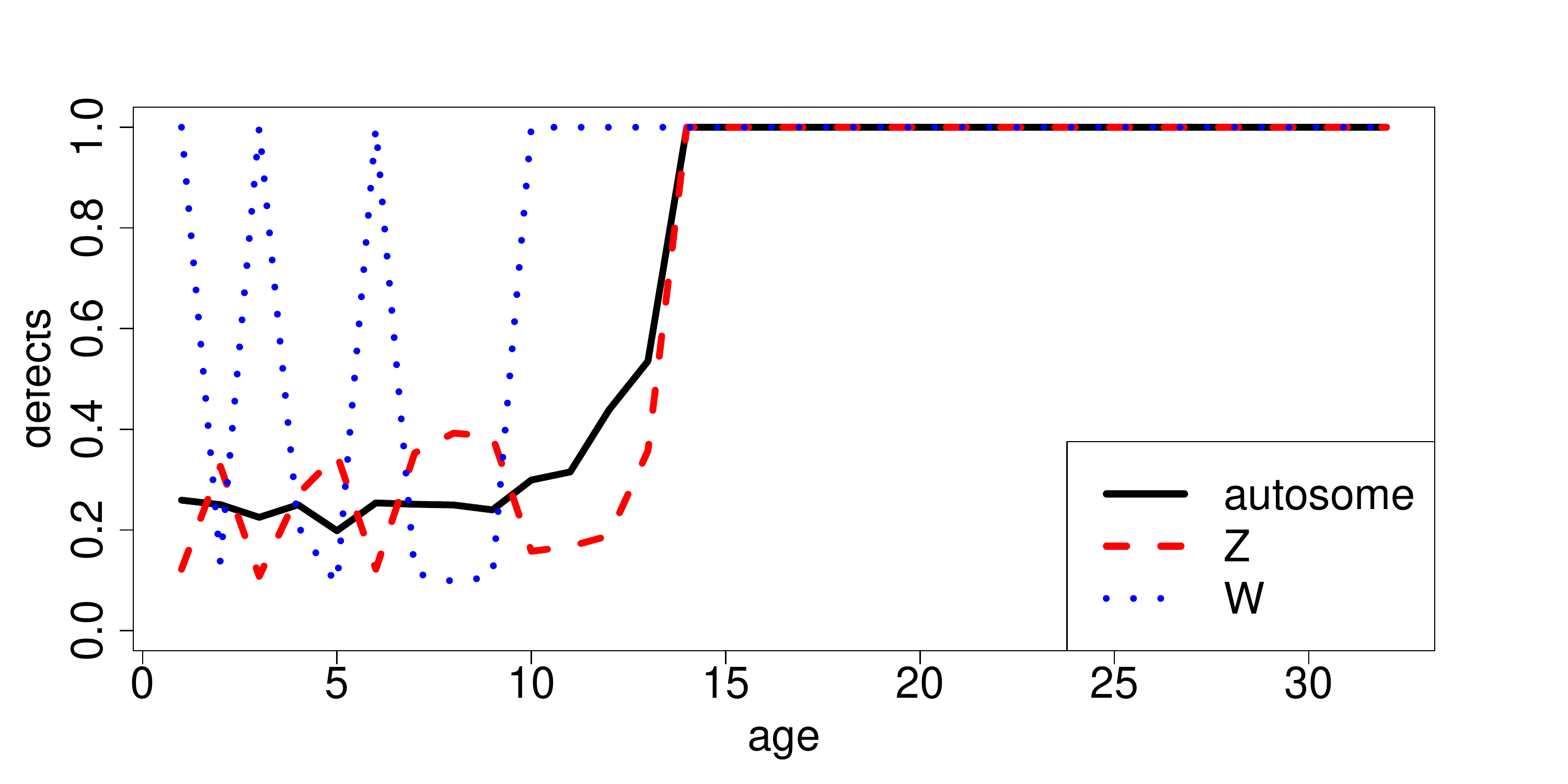}
}
	\caption{ZZ/WZ system - panmictic population. All descriptions like in
Fig. 1.}
\label{fig:6}       
\end{figure}

\begin{figure}
\resizebox{0.95\columnwidth}{!}{%
		\includegraphics[width=0.8\textwidth]{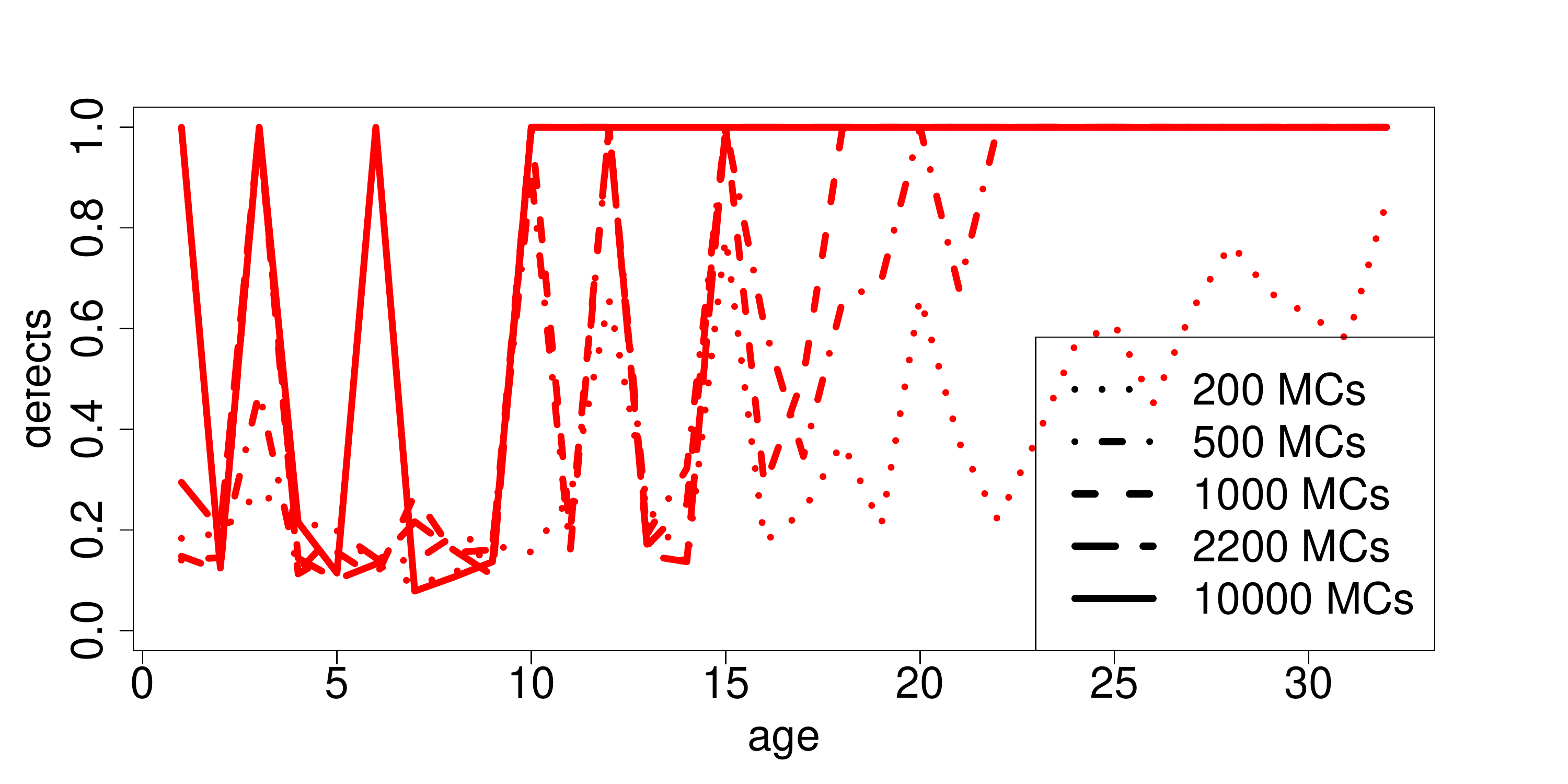}
}
\resizebox{0.95\columnwidth}{!}{%
		\includegraphics[width=0.8\textwidth]{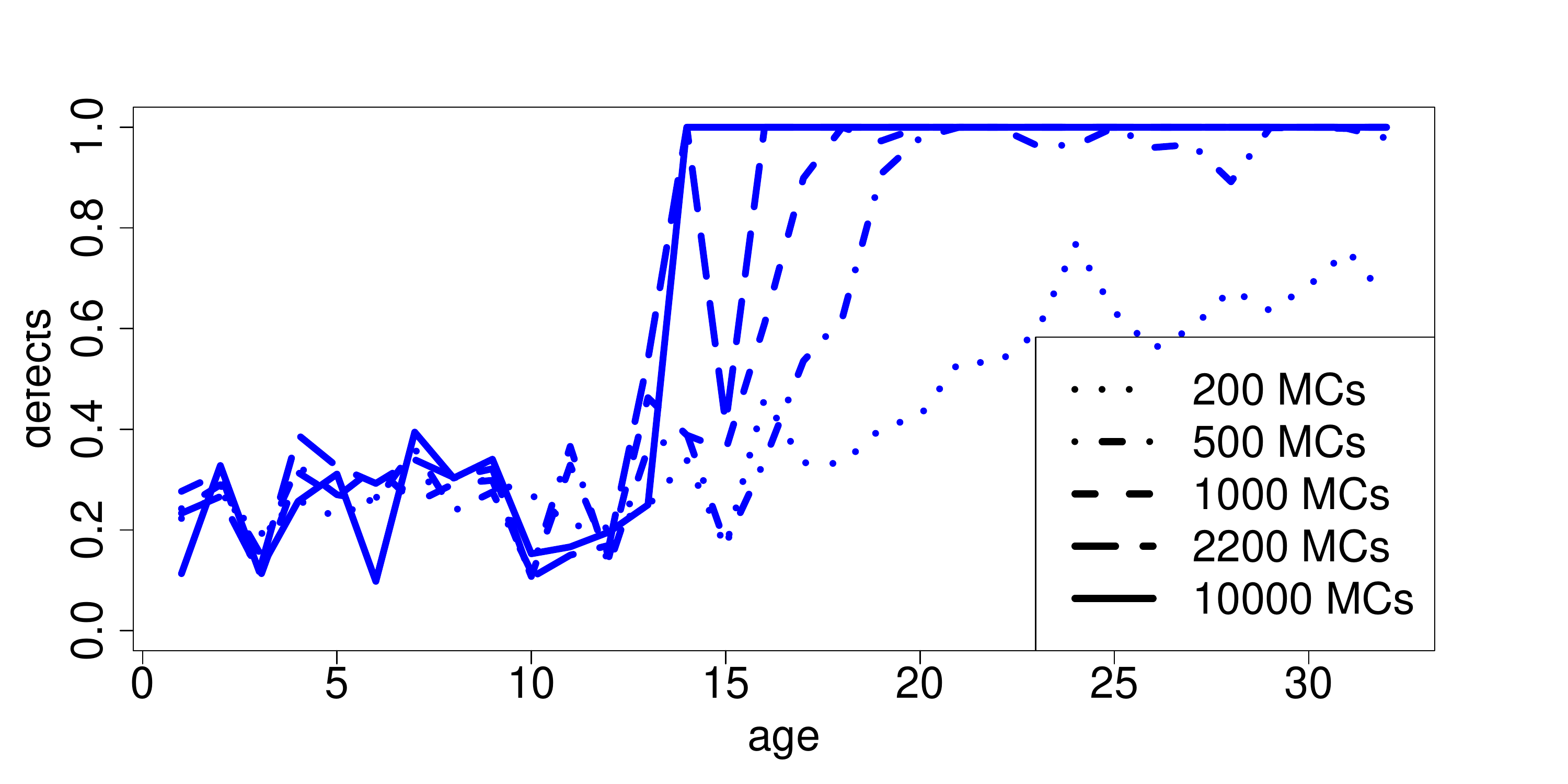}
}
	\caption{ZZ/WZ system - panmictic population. All descriptions like in
Fig. 2.}
\label{fig:7}       
\end{figure}

On the other hand, the male is also necessary during the whole period of hatching and feeding the offspring in the most cases of birds reproduction, even if eggs in the nest are not fertilized by him. The results of simulations (Fig. 4 and 5) show that the age distributions of both sexes are similar. We have observed variations in the fraction of defective genes along the sex chromosomes but higher fraction of defects on W chromosomes always corresponds to significantly lower fraction of defects in the same locus on Z chromosomes and vice versa. As a result, the age distributions and numbers of males and females in the population stay roughly the same. 

\section{Conclusion}
One can conclude that Y chromosome is shorter than X chromosome because females are promiscuous and seduce the males even if they  have been already successfully involved in the reproduction process with another female. Males have to pay for their feeble character with the higher mortality. It does not mean that all males have to be engaged into such unfaithful couples to reach the effect of shrinking Y chromosome. In our studies of hybridogenesis \cite{hybrids}, where temporal clonal reproduction of haplotypes can be observed (Y chromosome is clonally replicating), the influx of only a few percent of recombined haplotypes prevents the accumulation of defective alleles in the genomes. By comparing the effects, we can assume that even if a few percent of population uses the panmictic strategy, the effect of shrinking Y chromosomes should be observed. Our results confirm in some instances the results of the Onody group \cite{Lobo} but in some other points they are different. The Onody group modeled the panmictic population, that is why they observed the shrinking Y chromosome. Nevertheless, using the Penna model they observed the specific compensation of defective alleles placed on the X and Y chromosomes. Such an effect is characteristic for very low frequency of intragenomic recombination (here the recombination between X chromosomes in the process of the female gamete production) \cite{Zawierta}, \cite{Wojtek}. If the recombination is low, the whole genomes, not only the sex chromosomes tend to complement the defective alleles, especially in the highly inbreeding populations \cite{Bonkowska}. In fact, the Onody group switched off totally the recombination between the sex chromosomes in both homo- and heterogametic genomes. This forced the complementation and could also affect the results of the age distribution of both sexes. In the Heumann-H$\ddot{o}$tzel model \cite{Heumann}, modified by \cite{Medeiros} they observed the same relations in size between the males and females subpopulations as in our panmictic populations. Our results show that the effect of shrinking the chromosome determining the male heterogametic sex should be observed independently of the pair of homologous chromosomes. If one of them is marked by the sex determining gene (or group of genes) it has to shrink. One can think that it should enable easy transfer of sex determinants between pairs of chromosomes. Nevertheless, there is a problem of the dose compensation of information coded by the second sex chromosome in the female genome (inactivation of one of X chromosome in the female body cells \cite{Lyon}). The necessity of this mechanism can stabilize the sex chromosome pair.

Authors are grateful to D. Stauffer for discussion. This work was supported by the Polish State Committee for Scientific
Research, grant \# 105/E-344/SPB and Polish Foundation for Science. It was done in
the frame of European programs: COST Action P10 and NEST—GIACS.

\end{document}